\documentclass[12pt, preprint]{aastex}
\usepackage{graphicx}

\newcommand{\asec}{\ifmmode ^{\prime\prime}\else$^{\prime\prime}$\fi}
\newcommand{\etal}{et\,al.}

\newcommand{\grad}{\hbox{$^\circ$}}

\newcommand{\amin}{^{\prime}}

\newcommand{\lsim}{\!\!\!\phantom{\le}\smash{\buildrel{}\over
 {\lower2.5dd\hbox{$\buildrel{\lower2dd\hbox{$\displaystyle<$}}\over
                                 \sim$}}}\,\,}
\newcommand{\gsim}{\!\!\!\phantom{\ge}\smash{\buildrel{}\over
{\lower2.5dd\hbox{$\buildrel{\lower2dd\hbox{$\displaystyle>$}}\over
                               \sim$}}}\,\,}

\slugcomment{Accepted for publication in  ApJ Letters}
\shorttitle{A quasar pair in SDSS J1536+0441}
\shortauthors{Bondi \& P\'erez-Torres}

\begin{document}

\title{VLBI detection of an AGN pair in the binary black hole candidate 
SDSS J1536+0441}
\author{M. Bondi\altaffilmark{1} and  M-A. P\'erez-Torres\altaffilmark{2} } 
\altaffiltext{1}{INAF-Istituto di Radioastronomia, Via Gobetti 101, I-40129, 
Bologna, Italy}
\altaffiltext{2}{Instituto de Astrof\'{\i}sica de Andaluc\'{\i}a, CSIC, Apartado Correos
	3004, 18080 Granada, Spain}

\begin{abstract}
We present first pc-scale radio imaging of the radio-quiet candidate binary
black hole system SDSS J1536+0441. The observations were carried out by the
European VLBI Network at the frequency of 5 GHz and allowed to image SDSS
J1536+0441 with a resolution of $\sim 10$ mas ($\sim 50$ pc).
Two compact radio cores are detected at
the position of the kpc-scale components VLA-A and VLA-B, proving the
presence of two compact active nuclei with radio luminosity 
$L_R\sim 10^{40}$ erg s$^{-1}$, thus ruling out the possibility that the 
two radio sources are
both powered by one 0.1 pc binary black hole.
From a comparison with published 8.5 GHz flux densities we derived an 
estimate of the radio spectral index of the two pc-scale core. 
Both cores have flat or inverted spectral index, and at least for the case 
of VLA-A we can rule out the possibility that synchrotron self-absorption 
is responsible for the inverted radio spectrum. We suggest that thermal 
free-free emission from an  X-ray heated disk wind may be powering the 
radio emission in VLA-A.

\end{abstract}

\keywords{
quasars: individual (SDSS J153636.22+044127.0) --- galaxies: active --- 
radio continuum: general
}

\section{Introduction}
\label{Intro}

SDSS J153636.22+044127.0 (hereafter SDSS J1536+0441) is certainly a peculiar
and complex object. The attention to this low redshift QSO was drawn by
\citet[][hereafter BL09]{BL09} in a search for quasars exhibiting components 
with multiple
redshifts among the Sloan Digital Sky Survey \citep{Abaz09}. They found three
line systems at different redshifts: two sets of broad-line emission at
$z=0.3889$ and $z=0.3727$, and a third set of narrow absorption lines at
$z=0.3878$. Narrow emission lines were found associated only to the higher
redshift system. BL09 interpreted these peculiar features
in terms of a massive black hole binary (BHB) system within the same galaxy 
with separation of $\sim 0.1$ 
pc and masses of $M^{7.3}_\odot$ and $M^{8.9}_\odot$, each with its own
broad line region and sharing the same narrow line region. 
BHB systems are expected as the results of galaxy mergers, but few
compelling candidates have survived scrutiny. Furthermore, the so-called
``final parsec problem'' \citep{Bege80,MM01}  is not currently understood. 
In particular, dynamical
friction with the stellar background is ineffective in shrinking the binary
below separations smaller than 1 pc where gravitational radiation can complete
the coalescence of the two black holes within a Hubble time. 

After the BL09 publication, SDSS J1536+0441 was target of several
observations with the aim to confirm or dispute the BHB scenario.
\citet[][hereafter WL09]{WL09} used the VLA to image the quasar at 8.5 GHz 
and detected two radio
emitting components separated by 0.97$^{\prime\prime}$ (5.1 kpc)\footnote{We
assume a flat cosmology with $H_0=71$ km s$^{-1}$ Mpc$^{-1}$ and
$\Omega_m=0.27$.}. This suggests
the possibility that SDSS J1536+0441 is not a BHB system
separated by $\sim 0.1$ pc but a double quasar separated by $\sim 5$ kpc and
probably residing in a moderately rich cluster of galaxies. High
resolution ground based and HST observations \citep{Deca09a,Deca09b,LB09} 
indeed detected an optical counterpart to the secondary radio component but 
were not conclusive on the nature of the optical counterpart itself 
(obscured AGN or
elliptical galaxy) and on its relation with the peculiar optical spectrum.
The scenario of a superposition of two AGNs was dismissed by \citet{LB09}
and \citet{Chor10} on the basis that the regions responsible for the two
sets of broad lines are spatially coincident, even if some doubts remain
given that this result is based on observations carried out with a seeing
larger than the angular separation of the two objects along the slit
\citep{Deca09b}.

A third possible explanation was raised by \citet{Chor09,Chor10}. Based on
Palomar and Keck optical spectra they conclude that SDSS J1536+0441 is an
unusual member of the class of AGNs known as double-peaked emitters
\citep[e.g.][]{HF88,EH94,Stra03}. The same interpretation was given
independently by \citet{Gask10}. The main problem with this interpretation
is that there are striking differences between the Balmer-line profile in
SDSS J1536+0441 and all the other double-peaked emitters \citep{LB09}.
Finally, \citet{TG09} argued that SDSS J1536+0441 could be both a
double-peaked emitter and a BHB system.

All these possible scenarios, together with the ejected black hole hypothesis,
were critically discussed in \citet{LB09}.
No matter what is the real explanation for the observed properties of
SDSS J1536+0441, what we can surely say is that this is a puzzling and
possibly unique object.

In this Letter, we present for the first time high-resolution 
($\simeq 10$ mas corresponding to $\simeq 50$ pc) images
of the 5 GHz continuum radio emission of SDSS J1536+0441, obtained
with the European VLBI Network (EVN). These data 
detect pc-scale radio cores  both
in SDSS J1536+0441 and in the companion at $0.97\asec$ east of the quasar
proving the presence of two AGNs in this region.

\section{Pc-scale Imaging}
\label{Radio}

The European VLBI Network (EVN) was used to observe SDSS J1536+0441
at 5 GHz on 2009 October 23. 
The observations were carried at 1024 Mbit s$^{-1}$
sustained bit rate to exploit the large bandwidth capabilities of the EVN,
with an array which included all the EVN antennas. 
The pointing position was centered on the VLA-A component detected by 
WL09.
Observations were phase-referenced to the calibrator J1539+0430 with a duty
cicle of 5 minutes. The total time on source for SDSS J1536+0441 was about 5
hours. The strong and compact sources J1613+3412 and J0154+4743 were used as
fringe finders and bandpass calibrators.

The data were correlated and calibrated at the JIVE correlator.
Standard a priori gain calibration was performed using the measured
gains and system temperatures of each antenna.
The amplitude calibration was refined using the phase reference source.
Further data inspection, flagging and imaging were performed by the authors
using the NRAO AIPS software. No polarization or self-calibrations were
performed.

The $1\sigma$ r.m.s. noise in the final images is about 15 $\mu$Jy.

\section{Results and Discussion}
\label{Disc}
Figure 1 shows the field of SDSS J1536+0441 imaged with the EVN at 5 GHz
using natural weighting, which resulted in a beam of $12\times 7$ mas in 
position angle $10^\circ$ and a
$1\sigma$ r.m.s noise of 15 $\mu$Jy beam$^{-1}$.
Both radio sources, VLA-A and VLA-B following the notation used by WL09,
are clearly detected with signal-to-noise ratios (SNR) of 50 and 15, 
respectively. Contour plots of VLA-A and VLA-B are shown in Fig. 2.

\begin{figure*}
\includegraphics[width=15cm]{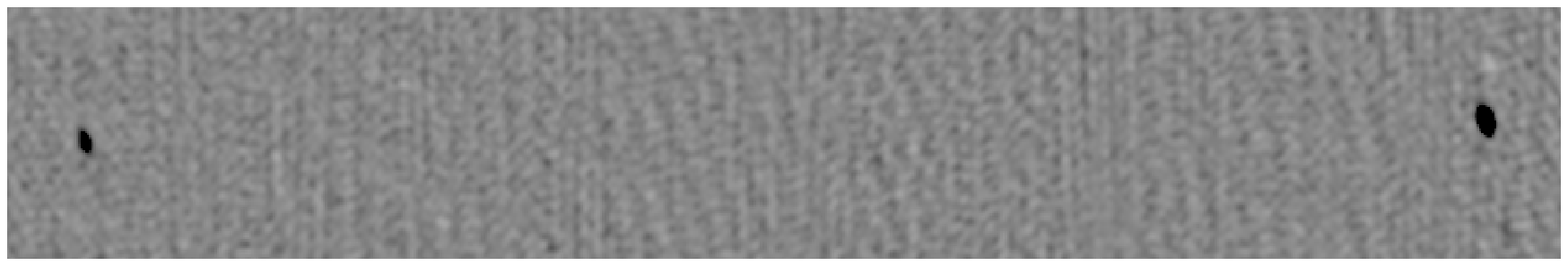}
\caption{EVN grey-scale image at 5 GHz of the SDSS 1536+0441 field. This
image was obtained applying natural weighting and has a resolution of
$12\times 7$ mas at position angle $10^\circ$ VLA-A (on the right) and VLA-B are clearly
distinguishable with SNR of 50 and 15 respectively.}
\label{fig1}
\end{figure*}

\begin{figure*}
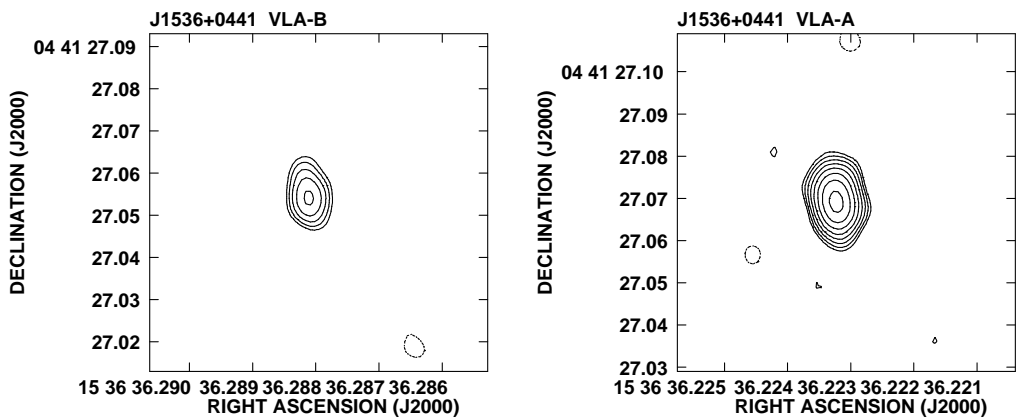

\includegraphics[width=6.5cm,angle=-90]{1536f2b.eps}
\includegraphics[width=6.5cm,angle=-90]{1536f2a.eps}
\caption{Contour plots of component VLA-A (right) and VLA-B (left) of  
SDSS 1536+0441. The beam size is $12\times 7$ mas at position angle 
$10^\circ$. The $1\sigma$ r.m.s noise is about 0.015 mJy beam$^{-1}$.
First contour is 3 times the noise and each inner contour is $\sqrt{2}$ times
brighter.
}
\label{fig2}
\end{figure*}

\placefigure{fig1}
\placefigure{fig2}

We also imaged the two radio sources with a slightly better resolution 
($7\times 6$ mas) but worse
sensitivity, using a different {\it u-v} data weighting function.
These images are not shown here but both set of images were used
to find consistent fitted Gaussian parameters for VLA-A and VLA-B.
The sources were fitted with two-dimensional elliptical Gaussians. 
The flux densities, positions and errors are:
for VLA-A, $S=0.72\pm 0.06$ mJy, 
$\alpha(J2000)=15^{\rm h}36^{\rm m}36^{\rm s}.2232$, 
$\delta(J2000)=+04\grad 41\amin 27\asec.069$, and $\sigma_{\rm VLBI}=0.003$
mas;
for VLA-B, $S=0.24\pm 0.03$ mJy,
$\alpha(J2000)=15^{\rm h}36^{\rm m}36^{\rm s}.2881$, 
$\delta(J2000)=+04\grad 41\amin 27\asec.054$, and $\sigma_{\rm VLBI}=0.003$
mas.
VLA-A appears slightly resolved when fitted with a single elliptical Gaussian 
component. The deconvolved fitted sizes are $\theta_{\rm VLA-A}\simeq
3.2\times 2.5$
mas with an estimated error of $1$ mas. The high SNR for this component
provides a high level of confidence for the deconvolved fitted size.
There is some indication that also VLA-B could be slightly resolved with
deconvolved size of about $2$ mas, but in this case given the limited SNR
this value should be considered as an upper limit.

The positions are in excellent agreement with those determined by WL09.
The separation between the two components is confirmed to be $0.97\asec$.

One of the goals of the proposed EVN observations was to determine the origin
of the radio emission in VLA-B and in particular if it could be associated
with VLA-A, e.g. as being a compact jet/hot-spot or a mini-lobe ejected by VLA-A.
No extended emission is detected in the region between the two compact radio
sources.
Given the compactness of VLA-B on the pc-scale
we can now confirm without any doubt that VLA-A and VLA-B are two compact
AGNs and we can rule out the possibility that the two radio sources are
both powered by a 0.1 pc binary (see WL09).

Comparing our flux densities at 5 GHz with those obtained by WL09 we
can derive an estimate of the radio spectral index between these two
frequencies.
It certainly can be speculative to draw conclusions on the radio spectral index
of VLA-A and VLA-B based on observations made at different resolutions and
epochs but it is a first step, waiting for more appropriate observations, and
it is illustrative of the general trend of the spectral properties.
As far as radio variability is concerned, the VLA 8.5 GHz and the VLBI 5 GHz
observations are separated by 8 months and significant variability 
(e.g. $\simeq 20\%$ with a few cases with larger variations)
is observed on such time range in radio-quiet and radio-loud
quasars \citep{Barv05}.
Keeping in mind this limitation, it is worth noting that the flux density of 
VLA-A at 5 GHz, 0.72 mJy, is 
lower than the 1.17 mJy measured at 8.5 GHz, while the flux density of VLA-B
is rather constant: 0.27 mJy at 8.5 GHz compared with 0.24 mJy at 5 GHz. 
Both sources would have flat or inverted radio spectra, and VLA-A in
particular would exhibit a strongly rising radio spectrum. 
WL09 already suggested
that VLA-A could have an inverted radio spectrum to explain the non detection
of the radio source in the 1.4 GHz FIRST sky survey \citep{Whit97} and our 5
GHz observations confirm this possibility. 
Another possibility is that we are missing flux in the 5 GHz VLBI
observations. If there is some extended emission (on the scale of $\simeq 100$
mas) our VLBI observations might not be deep enough or have the adequate  
{\it u-v} coverage to detect it. This possibility is rather improbable since
a significant amount of extended, and therefore steep spectrum emission at 5
GHz, 
should have been detected in the 1.4 GHz First survey given the 1 mJy 
threshold of the VLA survey.
Therefore, in the remaining discussion we will assume that any missing flux
is not significantly affecting the derived spectral indices.

With the measured flux densities at 5 and 8.5 GHz, VLA-A would have a
spectral index
$\alpha\simeq -0.9$ (with $S(\nu)\propto \nu^{-\alpha}$). Such an inverted
spectrum, even if found in a few radio quiet quasars, is quite unusual
\citep{Barv96,Kuku98,Ulve05}.
Synchrotron self-absorption is the most invoked cause to explain the flat or
inverted spectra in radio-loud or radio-quiet AGNs.
For self-absorption to occur, the brightness temperature must be comparable
to the kinetic temperature of the synchrotron electrons. The brightness
temperature in Kelvin is \citep[e.g.][]{Ulve05}:

\begin{equation}
T_b=1.8\times 10^9(1+z)\left(\frac{S_\nu}{1 {\rm mJy}}\right)\left(\frac{\nu}
{1 {\rm GHz}}\right)^{-2}\left(\frac{\theta_1 \theta_2}{1 {\rm mas}^2}\right)^{-1} 
\end{equation}

which for VLA-A gives $T_b=9\times 10^6$ K and for VLA-B $T_b\gsim 6\times
10^6$ K (assuming un upper limit for the size of VLA-B of $2\times 2$ mas).
These brightness temperatures, at least for VLA-A for which we have a
measured fitted size, are too
low to affect the spectra unless the magnetic fields are unrealistically high
\citep{Gall96}. 

The optical counterpart of VLA-B is very red \citep{Deca09b,LB09} and with the
detection of a pc-scale flat spectrum radio core with observed radio luminosity  
$L_R=\nu L_\nu$ at 5 GHz of $0.6\times 10^{40}$ erg s$^{-1}$ is best
interpreted as an obscured AGN rather than an elliptical galaxy. 
The brightness temperature of VLA-B is only a lower limit
since the size is not constrained and we cannot rule out the possibility of
synchrotron self-absorption affecting the flat radio spectrum as well as 
the contribution  of thermal free-free absorption/emission. 

The low brightness temperature of VLA-A rules out synchrotron self-absorption
as the origin of the inverted radio spectrum. Free-free absorption should
be discarded as well, since the view to the AGN is unobscured in VLA-A
where we see optical continuum emission from the AGN and the broad-line 
region.
We suggest that the radio emission from VLA-A could be interpreted as
thermal free-free emission from a disk wind \citep{Gall96,BK07}. 
The disk wind is heated by the X-ray continuum and therefore we can expect
a link between the radio and X-ray emission. 
\citet{Pane07} found a linear correlation between radio ($L_R$ at 5 GHz) 
and X-ray ($L_X$) core luminosities in an optically selected sample 
of nearby Seyfert. This result was confirmed and extended to radio-quiet
quasars by  \citet{LB08} using the carefully selected and almost complete
Palomar-Green quasar sample. They found that $L_R/L_X \sim 10^{-5}$ where 
$L_R$ is the radio luminosity at 5 GHz and $L_X$ is the bolometric 0.2-20 keV
X-ray luminosity. Quite remarkably, the same correlation, known as
G\"udel-Benz relation \citep{GB93}, holds for coronally active stars, 
therefore covering a range of about 15 orders of magnitude in luminosity.
WL09 used the X-ray luminosity measured by {\it Swift} \citep{Arzo09} 
and extrapolated the 5 GHz luminosity from their 8.5 GHz measurement
assuming a radio spectral index $\alpha=0.5$, obtaining $L_R/L_X=5.9\times
10^{-5}$.
Having a direct measure of the core flux density at 5 GHz we can refine this   
measurement. The observed monochromatic luminosity at 5 GHz of VLA-A is
$L_{\rm 5GHz}= 3.8\times 10^{30}$ erg s$^{-1}$ Hz$^{-1}$, and the radio
luminosity at 5 GHz is $L_R=\nu L_{\rm 5GHz}=1.9\times 10^{40}$ erg s$^{-1}$
which gives $L_R/L_X=1.4\times 10^{-5}$, assuming all the X-ray emission is
associated to VLA-A.

These VLBI observations were not meant to solve the so-far unsatisfactory 
interpretation of the puzzling properties of SDSS J1536+0441. The debate is
still open about all the possible scenarios summarised in the Introduction.
What is now clear is that both VLA-A and VLA-B are powered
by their own AGN, and this should be taken into account in future analysis.
Further radio observations are necessary to confirm the
spectral shape of the radio cores and therefore the origin of the 
radio emission.

\section{Summary}

We have presented the first VLBI pc-scale imaging of the candidate binary
black hole system
SDSS J1536+0441 observed by the European VLBI Network at 5 GHz. 
Both the VLA-A component, associated with SDSS J1536+0441, and the
companion object, VLA-B, at $0.97^{\prime\prime}$ east of the quasar are
detected with high SNR (50 and 15 respectively).
The two radio nuclei appears barely resolved and no extended larger scale 
emission is
detected. The main results can be summarised as follow:

\begin{itemize}
\item
We detect a flat spectrum pc-scale radio nucleus at the position of VLA-B
confirming that both VLA-A and VLA-B are powered by their own AGN.
Given the radio and optical properties, VLA-B is most likely associated to an
obscured AGN rather than a passive elliptical galaxy.

\item
At the position of VLA-A we detect a slightly resolved radio nucleus with a
strongly rising spectrum with $\alpha\simeq -0.9$ between 5 and 8.5 GHz. 
We rule out synchroron self-absorption as the cause of the inverted radio
spectrum because the derived brightness temperature of VLA-A is too low.
We suggest that thermal free-free emission from a disk wind provides the
simplest explanation for the inverted radio spectrum in VLA-A.

\item
We derive a value of $L_R/L_X=1.4\times 10^{-5}$ for VLA-A that is totally
consistent with correlation found for radio-quiet quasars and Seyfert 
galaxies \citep{Pane07,LB08}. 

\end{itemize}

\acknowledgments
MAPT acknowledges support by
the Spanish Ministry of Science and Innovation, and from the Andalusian
Research Ministry, through grants AYA2009-13036-C02-01 and TIC-126.
The European VLBI Network is a joint facility of European, Chinese,
South African and other astronomy institutes funded by their national
research councils. We wish to thank all the people at the telescopes
involved in the observations, those who took care to correlate and
process the data, and in particular Dr. Yurii Pidopryhora, support scientist
for these observations.

{\it Facility:} \facility{EVN}

\end{document}